\documentclass[12pt,a4paper]{article}
\usepackage{amsfonts}
\begin{document}
\begin{center}

{\LARGE{\bf{Exact Solution of the Quantum Calogero-Gaudin System }}}
\medskip
{\LARGE{\bf{and of its $q-$Deformation}}}

\vskip1cm

{\sc Fabio Musso  and Orlando Ragnisco$^\ast$ }
\vskip0.5cm

{\it  Dipartimento di Fisica, Universit\`a di Roma  TRE,}\\
{\it Via Vasca Navale 84, 00146 Roma, Italy} \\
{\it $^\ast$ I.N.F.N. - Sezione di Roma TRE, Roma, Italy}

\end{center}
\bigskip

\begin{abstract}
A complete set of commuting observables for the Calogero-Gaudin system 
is diagonalized, and the explicit form of the corresponding eigenvalues and 
eigenfunctions is derived. We use a purely algebraic procedure exploiting 
the co-algebra invariance of the model; with the proper technical 
modifications this procedure can be applied to the $q-$deformed version 
of the model, which is then also exactly solved.
\end{abstract}

\section*{Introduction}
In a number  of recent papers \cite{Rag},\cite{Ragn}, \cite{Herra} it has
 been pointed out
that co-algebras provide a simple and general mechanism to construct 
integrable Hamiltonian systems with an arbitrary number $N$ of degrees 
of freedom.  Moreover, as it relies upon 
the existence of a co-associative homomorphism, called ``co-product'' or
 ``co-multiplication'', from an algebra $\cal{A}$ to the
tensor product $\cal{A} \otimes \cal{A}$, this procedure works both in the
 standard ``Lie-algebra'' 
setting and in the so-called ``$q$-Lie-algebra'' setting.

In refs \cite{Rag},\cite{Ragn} the authors dealt mainly with classical 
Hamiltonian systems, where the Lie algebra, or its $q$-deformation, is
 realized 
in terms of Poisson brackets, but they have stressed that the same results,
``mutatis mutandis'' do hold for quantum systems as well. To avoid any possible
source of misunderstanding, from now on when using the word ``quantum''we  
will refer to the ``canonical'' Dirac quantization, while  in
the context of ``quantum groups'' or ``quantum algebras'' we will rather use  
the word ``deformed''.

The scope of this paper is to build up and solve a concrete example of 
a quantum integrable system
arising in the co-algebra setting, namely a  quantum version of the 
Calogero-Gaudin (CG) system \cite{C&vD}, \cite{Gaudin1}, \cite{Gaudin2},
both in the undeformed and in the deformed case.

Our starting point  will be the Calogero-Van Diejen paper \cite{C&vD}, where 
three integrable quantum hamiltonians related to CG have been considered.
It turns out that the relevant results can be formulated 
on pure algebraic grounds, without resorting to a specific realization:
the whole derivation will be then carried out in an abstract setting, 
holding for any infinite-dimensional representation; the Calogero-Van Diejen 
realization is recovered as a special case.  

Accordingly, in section 1 we 
construct the complete set of commuting observables for one of the 
Calogero-Van Diejen Hamiltonians and solve the associated spectral problem.

In section 2, we turn to the {\it{deformed}} quantum  system, and 
derive the spectrum and the common eigenfunctions for the corresponding set 
of observables.
On the way, we are naturally led to introduce what we have called 
``$z$-harmonic
polynomials'' (or ``$q$-harmonic polynomials'', or ``deformed harmonic 
polynomials'');
to the best of our knowledge, they are new mathematical objects, defined as 
polynomials solutions of a suitably deformed version of the $N$-dimensional
Laplacian.

In section 3, we mention some closely related open problems which, in our 
opinion,
deserve further investigation.

To speed up the presentation of the results, most of the details of the 
calculations 
for the {\it{undeformed}} (resp.{\it{ deformed}}) case are confined in 
Appendix I 
(resp. Appendix II).
  
\section{The quantum undeformed case}
\subsection{Calogero-Van Diejen results}

In \cite{C&vD} the authors discuss three different quantum versions of the 
classical CG, associated with the Hamiltonian function \cite{Calogero}:
\begin{equation}
H=\sum_{j,k=1}^N p_j p_k[ \lambda+ \mu \cos (q_j-q_k)] \label{classic}
\end{equation}
We will study the one which is associated with the Hamiltonian operator 
\begin{displaymath}
\hat{\mathcal{H}}=\lambda \hat{P}^2+\frac{\mu}{4}\hat{B}^+\hat{B} 
\end{displaymath}
\begin{eqnarray*}
&&\hat{P}=\sum_{j=1}^N \hat{p}_j \qquad \hat{p_j}= \frac{\hbar}{i} 
\frac{\partial}{\partial \hat{q}_j} \\
&&\hat{B}=\sum_{j=1}^N \hat{b}_j^2 \qquad \hat{b_j}=e^{-i \frac{\hat{q}_j}{2}}
 \sqrt{2 \hat{p}_j} 
\end{eqnarray*}  
The Hilbert space chosen by Calogero and Van Diejen is the subspace 
${\bf{H}}^{(+)}$ generated  by
\begin{equation}
\left\{ E_{\vec{m}}(\vec{q})=exp \left( i \sum_{j=1}^N m_j q_j \right) \quad
\vec{m} \in {\mathbb{N}}^{\it N } \right\} \label{basis}
\end{equation}
with inner product
\begin{displaymath}
(\phi,\psi)=\int \limits_0^{2 \pi} \ d q_1 \dots \int \limits_0^{2 \pi} 
\ d q_N \phi^*(\vec{q}) \psi(\vec{q})
\end{displaymath}
The domain of our operators will be the (linear) variety, everywhere dense in
${\bf{H}}^{(+)}$, of $C^\infty$ functions with Fourier components of 
nonnegative frequency, periodic of period $2 \pi$ in all
variables together with their derivatives. The choice of  ${\bf{H}}^{(+)}$ 
is an admissible one, as it is an 
invariant subspace for all the commuting observables.
As $[b_j,b_k^\dagger]=\hbar \delta_{jk}$, we may use (in units $\hbar=1$) 
the representation:
\begin{equation}
b_j^\dagger \longrightarrow x_j \qquad b_j \longrightarrow 
\frac{\partial}{\partial x_j}
\label{D}
\end{equation}
in which $\hat{\mathcal{H}}$ becomes
\begin{eqnarray*}
&&\hat{\mathcal{H}}_D=\frac{1}{4}(\lambda \hat{\cal{N}}^2+ \mu r^2 \nabla^2)
\qquad \hat{\cal{N}}=\sum_{j=1}^N \hat{\cal{N}}_j=\sum_{j=1}^N x_j 
\frac{\partial}{\partial x_j}\\
&&r^2=\sum_{j=1}^N x_j^2 \qquad 
\nabla^2=\sum_{j=1}^N \frac{\partial^2}{\partial x_j^2}  
\end{eqnarray*}
We note explicitly  that  in the  representation (\ref{D})  the operator 
$b^\dagger$ 
is nomore the
hermitian conjugate to  $b$, so that the Hamiltonian itself is nomore 
hermitian: however, the representation (\ref{D})  has been only used as 
an intermediate
technical step to  derive  the solutions, that at the end are  recast in the 
original variables, thus restoring hermiticity. 

The basis elements corresponding to (\ref{basis}) in representation 
(\ref{D}) are
the monomials: 
\begin{equation}
F_{\vec{m}}(\vec{x})=
\frac{x_1^{2m_1}}{\sqrt{2 m_1!}} \frac{x_2^{2m_2}}{\sqrt{2 m_2!}} \dots 
\frac{x_N^{2m_N}}{\sqrt{2 m_N!}} \qquad \vec{m} \in {\mathbb{N}}^N
 \label{monomials}
\end{equation}
We observe that the vacuum $|0 \rangle$
is given in both representations by a constant.  Acting iteratively on 
$|0 \rangle$ with the single-particle creation operator we obtain:
\begin{equation}
(b_j^\dagger)^l |0 \rangle=c x_j^l \longleftrightarrow c \sqrt{l!} 
\exp(iq_j l/2)
\label{substitution}
\end{equation}
Formula (\ref{substitution}) defines the (non-unitary) operator $\hat T$ 
intertwining between the two representations.

Calogero and Van Diejen have obtained the eigenvalues and eigenfunctions 
of the Hamiltonian in representation (\ref{D}):
\begin{displaymath}
\hat{\mathcal{H}}_D \phi_{k,m}(\vec{x})=\mathcal{E}_{km} \phi_{k,m}(\vec{x})
\end{displaymath}
where
\begin{enumerate}
\item  $\phi_{k,m}(\vec{x})=r^{2(k-m)} H_{2m}(\vec{x}) \ (k \geq m)$, 
$H_{2m}$ being
an even harmonic polynomial of degree $2m$ in $N$ variables
\item $\displaystyle \mathcal{E}_{km}=\lambda k^2+ 
\mu \left[ k^2+ \left( \frac{N}{2}-1 \right) k-m \left( m+\frac{N}{2}-1
 \right)\right]$
\end{enumerate}
The degeneracy of each eigenvalue is equal to the number of even independent 
harmonic 
polynomials of degree $2m$, i.e.
\begin{equation}
\left(
\begin{array}{c}
N-2+m\\
m
\end{array}
\right) \qquad N \geq 2 \label{binomio}
\end{equation}
Clearly, in the original representation the eigenvalues are the same, while
the eigenfunctions, that we denote by $\Psi_{k,m}(\vec{q})$
can be easily obtained from $\phi_{k,m}(\vec{x})$ through formula
(\ref{substitution}). Indeed, as $\phi_{k,m}(\vec{x})$ is an even homogeneous 
polynomial of degree $k$, it can be 
written in the form
\begin{displaymath}
\phi_{k,m}(\vec{x})=\sum_{l_1,\dots,\l_N=0 \atop l_1+\dots+l_N=k}^k c(\vec{m},
\vec{l})
\prod_{j=1}^N x_j^{2l_j} |0 \rangle
\end{displaymath}
where the coefficients 
$c(\vec{m},\vec{l})$ are determined by the particular 
choice for the basis of the harmonic polynomials.
Accordingly we have:
\begin{displaymath}
\Psi_{k,m}(\vec{q})=\sum_{l_1,\dots,\l_N=0 \atop l_1+\dots+l_N=k}^k
c(\vec{m},\vec{l})
\left( \prod_{j=1}^N (2l_j)! \right)^{1/2} \exp \left(i 
\sum_{j=1}^N l_j q_j \right)
\end{displaymath}

\subsection{Integrability of  Calogero-Van Diejen system}

Complete integrability of Calogero-Van Diejen system can be proved using 
an algebraic method to construct the integrals of motion. 
This method, first introduced by Karimipour \cite{Karimipour} 
while dealing with integrability of (\ref{classic}), has been later on 
cast in a more general setting in \cite{Rag}, \cite{Ragn}. 
The basic idea is the following:
suppose one is given a Poisson (resp. Lie) algebra $g$ realized by 
means of analytic 
functions of canonical phase space variables $(p,q)$ 
(resp. of canonical quantum
operators $(\hat p, \hat q)$) with Casimir ${\cal{C}} \in U(g)$, 
and a co-associative 
linear mapping $\Delta : U(g)\to U(g)\otimes U(g)$ (denoted as coproduct)
such that 
$\Delta$ is a Poisson (resp. Lie) homomorphism:

$$ [\Delta (a),\Delta (b)]_{U(g)\otimes U(g)}=\Delta ([a,b]_{U(g)})$$
It has been shown  \cite{Rag}, \cite{Ragn} that coassociativity allows one to 
construct from $\Delta$ in an unambiguous way subsequent homomorphisms 
$$\Delta^{(2)}:=\Delta, \quad \Delta^{(3)}:U(g)\to U(g)^{\otimes}{}^3,\dots ,
\Delta^{(N)}:U(g)\to U(g)^{\otimes}{}^N.$$
Thus, we can associate to our algebra, (or better {\it{co-algebra}}) a 
classical (resp. quantum) integrable system with $N$  degrees of freedom, 
whose Hamiltonian is an arbitrary (analytic) function of the $N^{th}$ 
coproduct 
of the generators and the remaining $N-1$ integrals of motion are provided by 
$\Delta^{(m)} ({\cal{C}})$\footnote{If the Hamiltonian turns out to be
functionally dependent upon the partial Casimirs $\Delta^{(m)}(C)$, a further
independent integral of motion is provided for instance by $\Delta^{(N)}(X)$,
$X$ being any of the generators.}, $m=2,\dots ,N$.
Incidentally, we notice here that Karimipour
stuck on the particular case where the coproduct is the one related to the 
usual Hopf-algebra structure defined on a universal enveloping algebra, namely:
$$\Delta (X) = X\otimes 1 + 1\otimes X \quad \forall \ X \in g$$  
Consider now the following $sl(2)$ realization in terms of $b$, $b^\dagger$
\begin{eqnarray}
\hat{X}^3&=&\frac{1}{2}(b^\dagger b + b b^\dagger) \nonumber\\
\hat{X}^+&=&\frac{(b^\dagger)^2}{2} \label{gener}\\ 
\hat{X}^-&=&-\frac{b^2}{2} \nonumber
\end{eqnarray}
The Quantum Casimir operators read
\begin{displaymath}
\hat{\mathcal{C}}_m=\frac{(\hat{X}^3_m)^2}{2}+\{\hat{X}^+_m,\hat{X}^-_m \}
\end{displaymath}
where, for a moment, we have used the notation 
$\hat{X}^i_m=\Delta^{(m)}(\hat{X}^i)$, 
$\hat{\mathcal{C}}_m=\Delta^{(m)}(\hat{\mathcal{C}})$.\\
The Hamiltonian can now be written as
\begin{displaymath}
\hat{\mathcal{H}}=\frac{1}{4}(\frac{\lambda}{2}+\mu)
(\hat{X}^3_N)^2-\frac{\mu}{2} 
\hat{\mathcal{C}}_N
+\frac{1}{32} \lambda
\end{displaymath}
So we have the following complete set of $N$ independent commuting operators:
\begin{equation}
\{\hat{\mathcal{C}}_2,\dots,\hat{\mathcal{C}}_N,\hat{\mathcal{H}} \} 
\label{family}
\end{equation}
\subsection{Solution to  the Spectral Problem}
In this section we will determine the spectrum for the complete set of 
commuting observables
\begin{equation}
\{\Delta^{(N)}(X_3),\Delta^{(2)}(\hat{\mathcal{C}}),
\dots,\Delta^{(N)}(\hat{\mathcal{C}}) \} 
\label{family2}
\end{equation}
As we said in the introduction, we will work, as far as possible, 
in a pure algebraic
setting. Accordingly, we will suppose that $\hat{X^+}$,  $\hat{X^-}$,  
$\hat{X^3}$,
are Hilbert space operators providing an infinite-dimensional representation of
 $sl(2)$.
Moreover, we will assume that the hermitian operator $\hat{X^3}$  is 
bounded from below, i.e., 
there exists a state $| 0 \rangle$ (called ``lowest weight vector'') such that 
\begin{displaymath}
X^3 | 0 \rangle=\lambda^{min} | 0 \rangle
\end{displaymath}
where $\lambda^{min}$ is the minimum in  the spectrum of $X^3$.

From $sl(2)$ commutation relations it follows that 
\begin{displaymath}
X^- | 0 \rangle=0
\end{displaymath}
Due  to the specific  form of the co-product, the 
lowest weight vector for $\Delta^{(n)}(X^3)$ will be simply the tensor product 
of the lowest weight vectors for the single particle operators 
$X^3_{(i)}$ \footnote{It might well happen that the kernel of $X^-$ is
not one-dimensional, implying that the different single particle spaces 
could be built up out of different ground states. This more general case 
has been thoroughly investigated in \cite{Harnad}.}:
\begin{displaymath}
\Delta^{(n)}(X^3) \overbrace{| 0 \rangle \cdots | 0 \rangle}^n=
\lambda^{min}_n   
\overbrace{| 0 \rangle \cdots | 0 \rangle}^n \qquad \lambda^{min}_n=
n \lambda^{min}
\qquad n=1,\dots,N
\end{displaymath}
As in the one-particle case commutation relations imply that, for each
$n$, the state $\overbrace{| 0 \rangle \cdots | 0 \rangle}^N$ belong to the
kernel 
of the operator $\Delta^{(n)}(X^-)$:
\begin{equation}
\Delta^{(n)}(X^-) \overbrace{| 0 \rangle \cdots | 0 \rangle}^N=0 
\qquad n=1,\dots,N
\label{ker}
\end{equation}
We will call $\overbrace{| 0 \rangle \cdots | 0 \rangle}^N$ the 
``ground state'' of the
system. 

Starting from the ground state, we will define the Hilbert space of the problem
as the space generated by the basis:
\begin{equation}
(X^+_1)^{n_1}(X^+_2)^{n_2} \cdots (X^+_N)^{n_N}
\overbrace{| 0 \rangle \cdots | 0 \rangle}^N \qquad n_i \in {\mathbb{N}} \ \ 
i=1,\dots,N \label{Hilbert}
\end{equation}
It is straightforward to see that substituting representation (\ref{gener}) 
in the
generators, one obtains exactly the same Hilbert space defined by Calogero and 
Van Diejen.  

The ground state turns out to be an eigenstate of {\it all} the Casimirs.
In fact, 
writing them in the form
\begin{displaymath}
\Delta^{(n)}({\cal{C}})=\frac{\Delta^{(n)}(X^3)^2}{2}-\Delta^{(n)}(X^3)+
\Delta^{(n)}(X^+)
\Delta^{(n)}(X^-)
\end{displaymath}
and using (\ref{ker}), we obtain:
\begin{equation}
\Delta^{(n)}({\cal{C}})\overbrace{| 0 \rangle \cdots | 0 \rangle}^N=
\lambda^{min}(\frac{\lambda^{min}}{2}-1)
\overbrace{| 0 \rangle \cdots | 0 \rangle}^N 
\label{lmc}
\end{equation}  
Now we are ready to prove the following proposition:
\newtheorem{Proposition}{Proposition}
\begin{Proposition}
The eigenfunctions of the complete set (\ref{family2}) of commuting 
observables are
of the form:
\begin{equation}
\phi_{k,m,s}=[\Delta^{(N)}(X^+)]^{k-m} H^{(2m)}_s \label{phi}
\end{equation}
where $H^{(2m)}_s$ is an ``s-particle harmonic polynomial'', i.e. it satisfies:
\begin{equation}
\Delta^{(s)}(X^-) H^{(2m)}_s=0 \label{harmonic}
\end{equation}
The harmonic polynomials are generated through the recursive formula:
\begin{eqnarray} 
&& H^{(2m)}_s= \left( \sum_{i=0}^{m-m'} a_{i,s,m,m^\prime} (X^+_{(s)})^{m-m'-i}
[\Delta^{(s-1)}(X^+)]^i \right) H^{(2m')}_{s'} \label{second} \\ 
&& \qquad m=1,2,\dots \qquad s=2,\dots,N \qquad m'<m,s'<s \nonumber\\
&& H^{(0)}_0=\overbrace{| 0 \rangle \cdots | 0 \rangle}^N
\end{eqnarray}  
where the constant $a_{i,s,m,m^\prime}$ must be chosen in such a way that 
(\ref{harmonic})
holds. 
\end{Proposition}
In formula (\ref{second}) we used the notation:
\begin{displaymath}
X^+_{(s)}=\overbrace{id \otimes \dots \otimes id}^{s-1} \otimes X^+ \otimes
\overbrace{id \otimes \dots \otimes id}^{n-s}
\end{displaymath}
{\bf{Proof:}} \vspace{12pt}\\
First of all we compute the commutator:
\begin{displaymath}
[\Delta^{(n)}(X^3),(X^+_{(s)})^{m-m'i}[\Delta^{(s-1)}(X^+)]^i] \qquad for 
\ n \geq s
\end{displaymath}
Now we write $\Delta^{(n)}(X^3)$ in the form:
\begin{eqnarray*}
&& \Delta^{(n)}(X^3)=\Delta^{(s-1)}(X^3) \otimes 
\overbrace{id \otimes \dots \otimes id}^{n-s+1}+\\
&&+ \overbrace{id \otimes \dots \otimes id}^{s-1} \otimes X^3 \otimes
\overbrace{id \otimes \dots \otimes id}^{n-s}+
\overbrace{id\otimes \dots \otimes id}^{s} \otimes \Delta^{(n-s)}(X^3),
\end{eqnarray*}
from which it follows:
\begin{eqnarray*}
&& [\Delta^{(n)}(X^3),(X^+_{(s)})^{m-m'-i}[\Delta^{(s-1)}(X^+)]^i]= 
(X^+_{(s)})^{m-m'-i}[\Delta^{(s-1)}(X^3),[\Delta^{(s-1)}(X^+)]^i]+\\
&&+[X^3_{(s)},(X^+_{(s)})^{m-m'-i}]\Delta^{(s-1)}(X^+)^i=\\
&& 2i (X^+_{(s)})^{m-m'-i}[\Delta^{(s-1)}(X^+)]^i+ 
2(m-m'-i) (X^+_{(s)})^{m-m'-i}[\Delta^{(s-1)}(X^+)]^i=\\
&& =2(m-m') (X^+_{(s)})^{m-m'-i}[\Delta^{(s-1)}(X^+)]^i
\end{eqnarray*} 
If we act with $\Delta^{(N)}(X^3)$ on an harmonic polynomial and we use 
repeatedly this last formula, we obtain:
 
\begin{displaymath}
\Delta^{(n)}(X^3) H^{(2m)}_s =(2m+\lambda^{min}_n)H^{(2m)}_s
\end{displaymath}
In particular, since $[\Delta^{(N)}(X^3),\Delta^{(N)}(X^+)]=
2 \Delta^{(N)}(X^+)$,
it follows that the functions $\phi_{k,m,s}$ (\ref{phi}) are eigenfunctions 
of the 
operator $\Delta^{(N)}(X^3)$ with eigenvalues given by:
\begin{displaymath}
\lambda_k= (2k+\lambda^{min}_N)
\end{displaymath}
Now we turn our attention to the remaining integrals of motions 
\begin{displaymath}
\{ \Delta^{(2)}(\hat{\mathcal{C}}), \dots ,\Delta^{(N)}(\hat{\mathcal{C}}) \} 
\end{displaymath}
First of all we note that from the homomorphism property of the coproduct it
follows that:
\begin{eqnarray*}
\Delta^{(n)}({\cal{C}})\phi_{k,m,s}&=&\Delta^{(n)}({\cal{C}})
\Delta^{(N)}(X^+)H^{(2m)}_s=\\
&=& \Delta^{(N)}(X^+) \Delta^{(n)}({\cal{C}}) H^{(2m)}_s \qquad n=2,\dots,N
\end{eqnarray*}
Hence we must worry only about the action of the partial Casimirs
on the harmonic polynomials. 
To this end we compute: 
\begin{displaymath}
\Delta^{(n)}({\cal{C}}) H^{(2m)}_s =\Delta^{(n)}({\cal{C}}) \left( \sum_{i=0}^{m-m'} a_{i,s,m,m^\prime} 
(X^+_{(s)})^{m-m'-i} [ \Delta^{(s-1)}(X^+) ]^i H^{(2m')}_{s'} \right)
\end{displaymath}
We distinguish two cases: if $n \geq s$, then condition (\ref{harmonic}) implies
\begin{eqnarray*}
\Delta^{(n)}({\cal{C}}) H^{(2m)}_s &=&\left( \frac{\Delta^{(n)}(X^3)^2}{2}-
\Delta^{(n)}(X^3) \right) H^{(2m)}_s=\\
&=& \left( \frac{(2m+\lambda^{min}_n)^2}{2}-
(2m+\lambda^{min}_n) \right) H^{(2m)}_s
\end{eqnarray*}
Viceversa, if $n<s$ then the Casimir operator 
$\Delta^{(n)}({\cal{C}})$ obviously 
commutes with the operator $X^+_{(s)}$; on the other hand, 
from the homomorphism 
property of the coproduct it follows that it commutes with the operator 
$\Delta^{(s-1)}(X^+)$ as well,
so that $\Delta^{(n)}({\cal{C}})$ acts directly on the harmonic polynomial
$H^{(2m')}_{s'}$
\begin{displaymath}
\Delta^{(n)}({\cal{C}}) H^{(2m)}_s= 
\sum_{i=0}^{m-m'} a_{i,s,m,m^\prime} (X^+_{(s)})^{m-m'-i}
[\Delta^{(s-1)}(X^+)]^i \Delta^{(n)}({\cal{C}}) H^{(2m')}_{s'} 
\end{displaymath}
Now we have again two possibilities: if $n \geq s'$ then we just showed that
$ H^{(2m')}_{s'}$ is eigenfunction of 
$\Delta^{(n)}({\cal{C}})$, hence $H^{(2m)}_s$
is an eigenfunction as well.

If $n<s$, then we will have:
\begin{displaymath}
\Delta^{(n)}({\cal{C}}) H^{(2m')}_{s'}= \sum_{i=0}^{m'-m''} 
a_{i,s,m',m''} (X^+_{(s)})^{m'-m''-i}
[\Delta^{(s-1)}(X^+)]^i \Delta^{(n)}({\cal{C}}) H^{(2m'')}_{s''} 
\end{displaymath}
We can iterate the above procedure until we reach the ground state $H^{(0)}_0$
or an harmonic polynomial $H^{(2m^{(i)})}_{s^{(i)}}$  
such that $n \geq s^{(i)}$. In both cases we know that it is eigenfunction
of the operator $\Delta^{(n)}({\cal{C}})$. Hence Proposition 1 is proved.
$\bullet$ \vspace{12pt}

Condition (\ref{harmonic}) implies the following recurrence relation for the
coefficients $a_{i, s, m,m^\prime}$:
\begin{eqnarray}
&& a_{i+1}=-\frac{(m-m'-i)[\lambda^{min}+m-m'-i-1]}{(i+1)[(s-1) \lambda^{min}
+i+2m']} a_i \label{rec2}\\
&& \qquad i=0,\dots,m-m'-1 \nonumber  
\end{eqnarray}
where for simplicity the labels $s,m,m^\prime$ have been omitted. 
Eq. (\ref{rec2}) can be easily ``solved'', yielding the following closed 
formula for the coefficients $a_{l, s, m,m^\prime}$:
\begin{equation}
a_{l}= (-1)^l\left(
\begin{array}{c}
m-m^\prime\\
l
\end{array}
\right){\frac{\Gamma(\lambda^{min} +m-m^\prime)}
{\Gamma(\lambda^{min} +m-m^\prime -l)}}
{\frac{\Gamma((s-1)\lambda^{min}+2(m^\prime -1))}
{\Gamma((s-1)\lambda^{min}+2(m^\prime -1)+l)}}a_0
\end{equation}
These results can be easily specialized to the realization (\ref{D})
used by Calogero and Van Diejen. First of all we note that the $sl(2)$
generators are expressed by:
\begin{eqnarray*}
&& X^+=\frac{x^2}{2}\\
&& X^-=-\frac{1}{2} \frac{\partial^2}{\partial x^2}\\
&& X^3= \frac{x}{2} \frac{\partial}{\partial x}+\frac{1}{4}
\end{eqnarray*}
and the corresponding coproducts by:
\begin{eqnarray*}
&& \Delta^{(n)}(X^+)=\frac{1}{2}\sum_{i=1}^n x^2_i\\
&& \Delta^{(n)}(X^-)=-\frac{1}{2} \sum_{i=1}^n 
\frac{\partial^2}{\partial x^2_i}\\
&& \Delta^{(n)}(X^3)= \frac{n}{4}  + \frac{1}{2} 
\sum_{i=1}^n x_i \frac{\partial}{\partial x_i}
\end{eqnarray*}
It follows that the ground state in this case is given simply by the 
constant function, with eigenvalue $\lambda^{min}_n=n/4$.
The polynomials (\ref{second}) are really harmonic (this is where the 
terminology comes from) and are given by the recursive formula:
\begin{eqnarray*}
&& H^{(0)}_0=c\\
&& H^{(2m)}_s=\left( \sum_{i=0}^{m-m'}a_{i,s,m,m'} r^{2i}_{(s-1)} 
x_s^{2(m-m'-i)}\right) H^{(2m')}_{s'} \\
&& s'=2,\dots,s-1 \qquad m'=1,\dots,m-1 \\
&& s=3,\dots,N \qquad m=2,3,4,\dots\dots 
\end{eqnarray*}
Actually, they form a basis in the space of harmonic polynomials.

\section{The quantum deformed case}
In \cite{Rag} it has been shown how to associate to a Poisson-Hopf
(Lie-Hopf) algebra a classical (quantum) integrable sistem and how to extend
this procedure to $q$-algebras. In fact, $q$-algebras are obtained by 
Poisson-Hopf algebras through a process of deformation that preserve their 
Poisson-Hopf structure. It is therefore possible to associate to $q$-algebras 
integrable systems that are deformed version of the ones associated
to the original algebra. 
Our aim is to analize the deformed version of the quantum system discussed in
section 1. The algebra to which this system is associated is $U(sl(2))$. The 
$q$-deformation of $U(sl(2))$, denoted by $U_q(sl(2))$, is well known
from the literature (see for example \cite{Hopf1}): the generators satisfy
the following commutation relations:
\begin{eqnarray}
&&[\tilde{X}^3,\tilde{X}^+]=2\tilde{X}^+ \nonumber\\ 
&&[\tilde{X}^3,\tilde{X}^-]=-2 \tilde{X}^- \label{commu}\\ 
&&[\tilde{X}^+,\tilde{X}^-]=\frac{\sinh(z\tilde{X}^3)}{\sinh z} \nonumber
\end{eqnarray}
and an admissible co-product is defined by:
\begin{eqnarray*}
&&\Delta(\tilde{X}^3)= \tilde{X}^3 \otimes 1 + 1 \otimes \tilde{X}^3 \\ 
&&\Delta(\tilde{X}^+)= \tilde{X}^+ \otimes e^{\frac{z\tilde{X}^3}{2}} +
 e^{-\frac{z\tilde{X}^3}{2}} \otimes \tilde{X}^+ \\ 
&&\Delta(\tilde{X}^-)= \tilde{X}^- \otimes e^{\frac{z\tilde{X}^3}{2}} + 
e^{-\frac{z\tilde{X}^3}{2}} \otimes \tilde{X}^-
\end{eqnarray*} 
(we prefer to use $z=\ln q$ as deformation parameter). 
We are going to realize this algebra in terms
of the operators $b$ and $b^\dagger$ introduced in section 1, in such a way 
that the
relation $(\tilde{X}^+)^\dagger=-\tilde{X}^-$ will hold whenever it does 
in the 
non-deformed case. We will see in a moment that this condition guarantees the 
hermiticity of the deformed Casimirs (\ref{qCasimir}).

A natural choice is to put:
\begin{eqnarray}
\tilde{X}^3&=&\frac{b b^\dagger +b^\dagger b}{2} \nonumber \\
\tilde{X}^+&=&f(z,\tilde{X}^3 -1)\frac{(b^\dagger)^2}{2} \label{choice}\\
\tilde{X}^-&=&-\frac{b^2}{2}f(z,\tilde{X}^3 -1) \nonumber
\end{eqnarray} 
$f(z,\tilde{X}^3)$ being an analitic function of the $\tilde{X}^3$ variable 
with a parametric dependence on $z$.
The realization (\ref{choice}) amounts to set to $-1/2$ the value of the 
one-body 
undeformed Casimir consistently with the ``bosonic'' realization 
(\ref{gener}). 

Imposing that these generators satisfy the commutation relations (\ref{commu})
we obtain a functional equation for $f(z,\tilde{X}^3)$ (see Appendix 2), a 
solution of which is given by:
\begin{equation}
f(z,\tilde{X}^3)=
\sqrt{\frac{4 \sinh^2 [z \tilde{X}^3 /2]+\sinh^2 z}{[(\tilde{X}^3)^2+1] 
\sinh^2 z}}
\label{f}
\end{equation}
We observe that, assuming the form (\ref{f}) for $f(z,\tilde{X}^3)$, we need 
invertibility of $(\tilde{X}^3-1)^2+1$ in (\ref{choice}), and this condition
is always verified if $\tilde{X}^3$ is hermitian.

The Casimir operator for this algebra is given by:
\begin{equation}
{\tilde{\cal{C}}}_z=\frac{1}{\sinh^2 z} 
\left\{ \sinh^2 \left[\frac{z(\tilde{X}^3+1)}{2}
\right]+\frac{}{} \sinh^2 \left[\frac{z(\tilde{X}^3-1)}{2} \right] \right\}
+\tilde{X}^+ \tilde{X}^-+ \tilde{X}^- \tilde{X}^+ \label{qCasimir}
\end{equation} 
It is easy to show that in the limit $z \longrightarrow 0$ we recover $sl(2)$
generators and Casimir.

Having the co-product and the Casimir, we can define an integrable quantum 
system with Hamiltonian ${\tilde{\mathcal{H}}}=\Delta^{(N)} 
({\tilde{\cal{C}}}_z)$
and integrals of motion given by $\Delta^{(N)}(\tilde{X}^3),\ 
\Delta^{(m)} ({\tilde{\cal{C}}}_z),\ m=2,\dots,N-1$ that is the 
$q$-deformation 
(actually the $z$-deformation) of the one treated in section 1; 
moreover, we can easily solve the associated spectral problem.

Indeed, using the commutation relations (\ref{commu}), the $n-$body Casimir
can be written in the following way:
\begin{displaymath}
\Delta^{(n)}({\tilde{\cal{C}}}_z)=2 \left( 
\frac{\sinh [z(\Delta^{(n)}(\tilde{X^3})-1)/2]}{\sinh z} \right)^2 +2 
\Delta^{(n)}(\tilde{X^+}) \Delta^{(n)}(\tilde{X^-})
\end{displaymath}
The crucial point is that, as in the undeformed case, the Casimir is the sum 
of a function of the coproduct of the $\tilde{X^3}$ generator plus the term 
$\Delta^{(n)}(\tilde{X^+}) \Delta^{(n)}(\tilde{X^-})$. This allows us to use 
the same procedure as in section $1$ to construct the eigenfunctions for the 
complete set of commuting observables:
\begin{equation}
\{ \Delta^{(N)}(\tilde{X}^-),\Delta^{(2)}({\tilde{\cal{C}}}_z),\dots,
\Delta^{(N)} ({\tilde{\cal{C}}}_z) \} \label{family3}
\end{equation}
We consider the same Hilbert space as defined in section 1. 
The lowest weight vector
($| 0 \rangle$) and its eigenvalue ($\lambda^{min}$) for the operator 
$\tilde{X^3}$ 
are the same as for $X^3$ since it is unchanged under deformation. 
Since the coproduct for the $\tilde{X^3}$ generator is itself unchanged,
the lowest weight vector for the operator $\Delta^{(N)}(\tilde{X^3})$ will be
again given by $\overbrace{| 0 \rangle \dots | 0 \rangle}^N$, 
which will be denoted
again as the ``ground state'' of the system. From commutation relations 
(\ref{commu}) follows that even in this case the ground state belongs to 
the kernel of the operators
$\Delta^{(n)}(\tilde{X^-}), \ n=2,\dots,N$.  

We have the following proposition:

\begin{Proposition}
The eigenfunctions of the complete set {\rm{(\ref{family3})}} of commuting 
observables are of the form:
\begin{displaymath}
\tilde{\phi}_{k,m,s}=[\Delta^{(N)}(\tilde{X^+})]^{k-m} \tilde{H}^{(2m)}_s
\end{displaymath}
where $H^{(2m)}_s$ is an ``s-particle deformed harmonic polynomial'', 
i.e. it satisfies:
\begin{equation}
\Delta^{(s)}(\tilde{X^-}) \tilde{H}^{(2m)}_s=0 \label{zharmonic}
\end{equation}
These deformed harmonic polynomials are generated through the recursive 
formula
\begin{eqnarray}
&& \tilde{H}^{(2m)}_s= \left( \sum_{i=0}^{m-m'} a_{i,s,m,m'}(z)
(\tilde{X}^+_{(s)})^{m-m'-i}[\Delta^{(s-1)}(\tilde{X^+})]^i \right)
\tilde{H}^{(2m')}_{s'} \label{qsecond} \\ 
&& m=1,2,\dots \qquad s=2,\dots,N \qquad m'<m,s'<s \nonumber\\
&& \tilde{H}^{(0)}_0=\overbrace{|0 \rangle \cdots |0 \rangle}^N \nonumber
\end{eqnarray}  
where the functions $a_{i,s,m,m'}(z)$ must be chosen in such a way that 
(\ref{zharmonic}) holds.
\end{Proposition}
In proposition $2$ we used the notation:
\begin{displaymath}
\tilde{X}^+_{(s)}=\overbrace{id \otimes \dots \otimes id}^{s-1} 
\otimes \tilde{X}^+ \otimes
\overbrace{id \otimes \dots \otimes id}^{n-s}
\end{displaymath}

The proof of this proposition proceeds in the same way as in the 
undeformed case, with some minor changes.
The eigenvalues of the partial Casimirs corresponding to the ground state are 
defined by the formula:
\begin{displaymath}
\Delta^{(n)} ({\tilde{\cal{C}}}_z) \overbrace{| 0 \rangle \dots | 0 \rangle}^N=
2 \left( \frac{\sinh [z(\lambda^{min}_n-1)/2]}{\sinh z} \right)^2
\overbrace{| 0 \rangle \dots | 0 \rangle}^N
\end{displaymath} 
On the other hand, on a generic excited state, corresponding to a 
deformed harmonic 
polynomial $\tilde{H}^{(2m)}_s$ we have (for $n \geq s$):
\begin{displaymath}
\Delta^{(n)} ({\tilde{\cal{C}}}_z) \tilde{H}^{(2m)}_s=
2 \left( \frac{\sinh [z(2m + \lambda^{min}_n-1)/2]}{\sinh z} \right)^2
\tilde{H}^{(2m)}_s 
\end{displaymath}   
The recurrence relation for the coefficients $a_{i,s,m,m^\prime}(z)$ 
is given by:
\begin{eqnarray*}
a_{i+1}(z)=-a_i(z)e^{\frac{z}{2}[2(m'+i-1)+(s-2) \lambda_{min}]}
\frac{  \sinh[z(\lambda_{min}+m-m^\prime -i-1)]\sinh[z(m-m^\prime -i)]}{
\sinh[z(2m'+\lambda_{min}(s-1)+i)]\sinh[(z(i+1)]}\\
\end{eqnarray*}
which is manifestly the $z-deformed$ version of the recurrence relation 
(\ref{rec2}).

\section{Concluding remarks}

As it is well known, the Calogero-Gaudin system is superintegrable, both at 
the classical and at the quantum level. An algebraic explanation for that 
property has been recently proposed by Ballesteros et al. \cite{PBH},
in the context of the ``two-photon algebra". An alternative interpretation 
relies on the fact that the ``two-body Casimirs" ${\cal{C}}_2 ^{(ij)}:= 
(\Delta^{(2)}({\cal{C}}))_{i,j}$, which 
(Poisson) commute with $\Delta ^{(N)}({\cal{C}})$, are actually the squares 
of the generators of $SO(N)$; this readily entails that the quantities:

$$ I_j = \sum_{k \ne j} 
\frac{{\cal{C}}_2 ^{(ij)}}{\lambda_j -\lambda_k}~~~~~~k=1,
\dots,N;~~\sum_k I_k = 0$$

\noindent
commute in pairs for any choice of the (distinct) numbers $ \{\lambda _j\}$.

The quantum system characterized by $\{\Delta ^{(N)}({\cal{C}}), I_j\}$ 
has been extensively studied in the recent past for finite dimensional 
representations of $sl(2)$ \cite{Gaudin1}, \cite{Gaudin2}, \cite{Sklianin}
through Bethe Ansatz and/or Quantum Inverse Scattering Method; 
for an infinite-dimensional representation, we refer again to \cite{Harnad}.

What about superintegrability of the $q-$deformed system? So far, no 
$q-$deformed analog of the family $\{\Delta ^{(N)}({\cal{C}}), I_j\}$ 
has been found \cite{Nostro}, and moreover a strong -though not 
compelling - no-go argument 
has been recently raised in \cite{PBH2}, based on the underlying 
$r-$matrix structure. We are actively working on this point to achieve  a 
definite answer.

A further open issue is the solution of the spectral problem for the 
quantum deformed CG model in a finite-dimensional representation of $sl_q(2)$. 
Work is in progress on that, and we expect to get the results shortly: 
indeed, due to its  purely algebraic nature,  the approach we have followed 
here can be applied with the proper technical modifications to the 
finite-dimensional case as well.
  
\section*{Appendix 1}
In this appendix we want to show that the
set of eigenfunctions $\phi_{k,m,s}$ (\ref{phi}) form a
basis respect to the Hilbert space of the problem. 

To this aim we give the following proposition
\begin{Proposition}
The total number of polynomials $H^{(2m)}_s$ (\ref{second}) of fixed degree
with $s=2,\dots,N$ is given by:
\begin{equation}
h(2m)_N=
\left(
\begin{array}{c}
N-2+m\\
m
\end{array}
\right) \label{number}
\end{equation}
\end{Proposition}
{\bfseries Proof:}\vspace{8pt}\\
If $s=2$ we can apply our recursive formula (\ref{second}) to the state 
$H^{(0)}_0$  so that we can construct 
only one harmonic polynomial for each value of $m$, i.e. $h(2m,2)=1$ . 
For $s=3$ and $m$ fixed, the recursive formula (\ref{second}) can be applied
either to $H^{(0)}_0$ or to a two particle harmonic polynomial with $m'<m$, 
so that
\begin{displaymath}
h(2m,3)=1+\sum_{i_1=1}^{m-1} 1
\end{displaymath}
Following this line of reasoning it is clear that, given $m$, for a
generic $s$ we have
\begin{eqnarray*}
h(2m,s)&=&1+\sum_{i_{s-2}=1}^{m-1}\left( 1+\sum_{i_{s-3}=1}^{i_{s-2}} \left(
\dots \left(1+\sum_{i_2=1}^{i_3}\left( 1+\sum_{i_1=1}^{i_2} 1\right)
\right) \dots \right) \right)=\\
&=&1+\sum_{i_{s-2}=1}^{m-1} 1 + \sum_{i_{s-2}=1}^{m-1} \sum_{i_{s-3}=1}
^{i_{s-2}} 1+\dots \dots+ \sum_{i_{s-2}=1}^{m-1} \dots \sum_{i_1=1}^{i_2} 1
\end{eqnarray*}
We want now to prove that, for $s>2$, it holds:
\begin{displaymath}
\sum_{i_{s-2}=1}^{m-1} \dots \sum_{i_1=1}^{i_2} 1=\frac{(s+m-4)!}{(s-2)!(m-2)!}
\end{displaymath}
We use induction. For $s=3$ the claim is trivial. Assuming the claim to hold 
for $s-1$,
for $s$ we have:
\begin{displaymath}
\sum_{i_{s-2}=1}^{m-1} \dots \sum_{i_1=1}^{i_2} 1=\sum_{i_{s-2}=1}^{m-1}
\frac{(s+i_{s-2}-4)!}{(s-3)!(i_{s-2}-1)!}
\end{displaymath}
From \cite{librus} we know that
\begin{equation}
\sum_{k=1}^n \frac{(k+m)!}{(k-1)!}=\frac{1}{(m+2)} \frac{(n+m+1)!}{(n-1)!}
\label{lr}
\end{equation}
In our case it means
\begin{eqnarray*}
\sum_{i_{s-2}=1}^{m-1}\frac{(s+i_{s-2}-4)!}{(s-3)!(i_{s-2}-1)!}=
\frac{1}{(s-3)!} \sum_{i_{s-2}=1}^{m-1} \frac{[(s-4)+i_{s-2})]!}
{(i_{s-2}-1)!}=\\
=\frac{1}{(s-3)!} \frac{1}{(s-2)} \frac{(s+m-4)!}{(m-2)!}=
\frac{1}{(s-2)!} \frac{(s+m-4)!}{(m-2)!}
\end{eqnarray*}
that proves our claim.

We can write $h(2m,s)$ in the form:
\begin{displaymath}
h(2m,s)=\sum_{i=2}^s \frac{(i+m-4)!}{(i-2)!(m-2)!}
\end{displaymath}
so that the total number of harmonic polynomials of degree $2m$ in $N$ 
variables is given by:
\begin{displaymath}
h(2m)_N=\sum_{s=2}^N \sum_{i=2}^s \frac{(i+m-4)!}{(i-2)!(m-2)!}
\end{displaymath}
Rescaling the indices and repeatedely using (\ref{lr}), we have:
\begin{eqnarray*}
h(2m)_N&=&\sum_{s=2}^N \sum_{i=2}^s \frac{(i+m-4)!}{(i-2)!(m-2)!}=
\sum_{s=1}^{N-1} \sum_{i=1}^s \frac{(i+m-3)!}{(i-1)!(m-2)!}=\\
&=&\sum_{s=1}^{N-1} \frac{1}{(m-1)!} \frac{(s+m-2)!}{(s-1)!}=
\frac{(m+N-2)!}{(m)!(N-2)!}
\end{eqnarray*}
that proves the proposition. $\bullet$ \vspace{12pt} 

We want now to show that our eigenfunctions form a basis for the 
Hilbert space of
the problem. We recall that the Hilbert space was generated by the monomials
(\ref{Hilbert}). We can decompose this space as the direct sum of the spaces of
homogeneous polynomials of degree $m$ for $m=0,
\dots,\infty$ that we denote with $P^{(N)}_{m}$. A basis in $P^{(N)}_{m}$
is obviously given by the monomials 
\begin{eqnarray*}
&& (X^+_1)^{n_1} (X^+_2)^{n_2} \dots (X^+_N)^{n_N}
\overbrace{|0 \rangle \dots |0 \rangle}^N \\
&& n_i=0,\dots,m \qquad i=1,\dots,N \qquad \sum_{i=1}^N n_i=m
\end{eqnarray*}
It follows that the dimension of $P^{(N)}_{m}$ space is given by 
\begin{equation}
{\rm{dim}} \, P^{(N)}_{m}=\left(
\begin{array}{c}
N+m-1\\
m
\end{array}
\right) \label{dimension}
\end{equation} 
We claim that this is also the number of the eigenfunctions $\phi_{k,m,s}$
of the form:
\begin{equation}
\phi_{k,m,s}=[\Delta^{(N)}(X^+)]^{k-m} H^{(2m)}_s \label{poly}
\end{equation}
In fact, given $k$, the number of harmonic polynomials is given by
\begin{displaymath}
\frac{(N+k-2)!}{k!(N-2)!}
\end{displaymath}
Hence the total number of polynomials of the form (\ref{poly}) is given by
\begin{displaymath}
\frac{1}{(N-2)!} \sum_{k=0}^m \frac{(N+k-2)!}{k!}
\end{displaymath}
Using again (\ref{lr}) we have that:
\begin{eqnarray*}
&&\frac{1}{(N-2)!} \sum_{k=0}^m \frac{(N+k-2)!}{k!}=\frac{1}{(N-2)!}
\sum_{k=1}^{m+1} \frac{(N+k-3)!}{(k-1)!}=\\
&&=\frac{1}{(N-2)!} \frac{1}{(N-1)} \frac{(N+m-1)!}{m!}={\rm{\dim}} 
\, P^{(N)}_{2m}
\end{eqnarray*}
that proves our claim.

\section*{Appendix 2}
In this appendix we derive in detail formula (\ref{f}).
First of all we observe that with the choice (\ref{choice}) the 
first two commutation
rules in (\ref{commu}) are automatically satisfied, while the third give us a 
functional equation for $f(z,\tilde{X}^3)$. We pose for shortness:
\begin{displaymath}
x=\frac{(b^\dagger)^2}{2} \qquad y=-\frac{b^2}{2} \qquad [x,y]=\tilde{X}^3 
\end{displaymath}                                                             
So that 
\begin{displaymath}
\tilde{X}^+=f(z,\tilde{X}^3-1)x \qquad \tilde{X}^-=yf(z,\tilde{X}^3-1)
\end{displaymath}
The commutator between $\tilde{X}^+$ and $\tilde{X}^-$ is hence given by
\begin{eqnarray}
[\tilde{X}^+,\tilde{X}^-]&=&f(z,\tilde{X}^3-1)y[x,f(z,\tilde{X}^3-1)]
+f^2(z,\tilde{X}^3-1) \tilde{X}^3+ \nonumber \\
&+&[f(z,\tilde{X}^3-1),y]f(z,\tilde{X}^3-1)x
\label{commX}
\end{eqnarray}
The commutator between $x,y$ and whatever analytic function 
$f(\tilde{X}^3)$ is 
given by:
\begin{eqnarray}
[f(\tilde{X}^3),x]=[f(\tilde{X}^3)-f(\tilde{X}^3-2)]x \label{x1}\\
\left[ f(\tilde{X}^3),y \right]= \left[ f(\tilde{X}^3)-f(\tilde{X}^3+2) 
\right]y \label{y1}
\end{eqnarray}             
Using the equations (\ref{x1}), (\ref{y1}), formula (\ref{commX}) becomes:
\begin{displaymath}
[\tilde{X}^+,\tilde{X}^-]=f^2(z,\tilde{X}^3-1) \tilde{X}^3+
[f^2(z,\tilde{X}^3-1)-f^2(z,\tilde{X}^3+1)]yx
\end{displaymath}
The product $yx$ as a function of $\tilde{X}^3$ reads:
\begin{displaymath}
yx=-\frac{1}{4}[(\tilde{X}^3+1)^2+1]
\end{displaymath}
So, finally, we have
\begin{equation}
[\tilde{X}^+,\tilde{X}^-]=\frac{1}{4} 
\{ f^2(z,\tilde{X}^3+1)[(\tilde{X}^3+1)^2+1]-
f^2(z,\tilde{X}^3-1)[(\tilde{X}^3-1)^2+1] \} \label{eq1}
\end{equation}
Imposing
\begin{equation}
F(z,\tilde{X}^3)=f^2(z,\tilde{X}^3)[(\tilde{X}^3)^2+1] \label{sostitu}
\end{equation}
equation (\ref{eq1}) becomes:
\begin{displaymath}
[\tilde{X}^+,\tilde{X}^-]=\frac{1}{4}[F(z,\tilde{X}^3+1)-F(z,\tilde{X}^3-1)]
\end{displaymath}
The requirement 
\begin{displaymath}
[\tilde{X}^+,\tilde{X}^-]=\frac{\sinh (z\tilde{X}^3)}{\sinh z}
\end{displaymath}
entails the following functional equation for $F(z,\tilde{X}^3)$:
\begin{equation}
\frac{1}{4}[F(z,\tilde{X}^3+1)-F(z,\tilde{X}^3-1)]
=\frac{\sinh(z\tilde{X}^3)}{\sinh z} \label{eq2}
\end{equation}
A solution of (\ref{eq2}) is given by:
\begin{equation}
F(z,\tilde{X}^3)=\frac{2 \cosh(z\tilde{X}^3)}{\sinh^2 z}+\rho(z) \label{solu}
\end{equation}
where $\rho(z)$ is an arbitrary function of $z$.

If now we substitute (\ref{solu}) in (\ref{sostitu}) we obtain an algebraic
equation for $f(z,\tilde{X}^3)$:
\begin{displaymath}
\frac{2 \cosh(z \tilde{X}^3)}{\sinh^2 z}+\rho(z)=
f^2(z,\tilde{X}^3)[(\tilde{X}^3)^2+1]
\end{displaymath}
This is an underdetermined equation, but we have yet to require that in the
limit $z \rightarrow 0$ the generators
$\tilde{X}^3,\tilde{X}^+,\tilde{X}^-$ must reproduce the
non-deformed ones (\ref{gener}). This means
\begin{displaymath}
{\rm{lim}}_{z \rightarrow 0} f(z,\tilde{X}^3)=1
\end{displaymath}
so that 
\begin{displaymath}
\rho(z)_{z \rightarrow 0} \sim -\frac{2}{z^2}+1
\end{displaymath}
An handy choice for $\rho(z)$ that possess this behaviour is given by:
\begin{displaymath}
\rho(z)=-\frac{2}{\sinh^2 z}+1
\end{displaymath}
from which it follows 
\begin{displaymath}
F(z,\tilde{X}^3)=\frac{4 \sinh^2(z\tilde{X}^3/2)}{\sinh^2 z}+1
=f^2(z,\tilde{X}^3)[(\tilde{X}^3)^2+1]
\end{displaymath}
which yields formula (\ref{f}).

\end{document}